\documentstyle[12pt]{article}
\begin{document}
\begin{titlepage}
\begin{flushright}
UM-TH-95-13\\
April 1995\\
\end{flushright}
\vskip 2cm
\begin{center}
{\large\bf Unstable Particles\footnote{Invited talk presented at
{\sl Perspectives for Electroweak Interactions in $e^+e^-$ Collisions},
Ringberg Castle, Tegernsee, Germany, 5--8 Februay, 1995.}
}
\vskip 1cm
{\large Robin G. Stuart}
\vskip 1cm
{\it Randall Physics Laboratory,\\
 University of Michigan,\\
 Ann Arbor, MI 48190-1120,\\
 USA\\}
\bigskip
and\\
\bigskip
{\it Instituto de F\'\i sica,\\
 Universidad Nacional Aut\'onoma de M\'exico,\\
 Apartado Postal 20-364, 01000 M\'exico D. F.}\\
\end{center}
\vskip .5cm
\begin{abstract}
Unstable particles cannot be treated as asymptotic external states in
$S$-matrix theory and when they occur as resonant states cannot be
described by finite-order perturbation theory.
The known facts concerning unstable particles are reviewed and it is
shown how to construct gauge-invariant expressions for matrix elements
containing intermediate unstable particles and physically meaningful
production cross-sections for unstable particles. The results
and methodology presented
are relevant for $Z^0$ resonance physics, $W^+W^-$ and $Z^0Z^0$ pair
production and can be straightforwardly applied to other processes.
\end{abstract}

\end{titlepage}

\setcounter{footnote}{0}
\setcounter{page}{2}
\setcounter{section}{0}
\newpage

\section{Introduction}

It is a fact of life that the heaviest known elementary particles,
the $W$ and $Z^0$ bosons and the top quark have total decay widths
that are a substantial fraction of their masses.
Yet $S$-matrix theory does not easily deal with unstable particles
as they cannot be represented by asymptotic states.
The study of width effects is timely and important for the theoretical
understanding of current and future experimental results from SLC, LEP
and hadron colliders.
The unstable $Z^0$ boson has by now been produced in vast numbers
at SLC and LEP and the latter machine is soon begin producing
$W^+W^-$ pairs.  Width effects have been suggested
as a possible mechanism for generating enhanced CP-violating
observables that could provide a window into new
physics.\cite{Pilaftsis,EilaHeweSoni} A number of attempts
have been made to treat width effects by introducing an imaginary part
by hand into the propagator of the unstable
particle\cite{Moriond,NowaPila,BaurZeppenfeld,Papadopoulos}.
Such approaches wreak havoc with
unitarity and current conservation and must then be patched in some
manner. In patching it is, in some cases, found necessary
to introduce an imaginary part into propagators of unstable particles
in the $t$-channel which is obviously incorrect.
It is not easy to extend such methods to
higher orders. A comparison of the relative merits of various approaches
appears in ref.s~\cite{AeppCuypOlde} and \cite{AeppOldeWyl}.
The analysis presented here suffers from none of these
difficulties and is fully consistent with constraints imposed by
analytic $S$-matrix theory and perturbation theory. Much of what appears
here can be found in ref.s~\cite{Stuart1,Stuart2,Stuart3} and
\cite{Stuart4}.

In this paper we consider two distinct reactions involving unstable
particles. The first is $e^+e^-\rightarrow f\bar f$ near the $Z^0$
resonance with $f$ used to denote a generic light fermion species.
This reaction proceeds via two distinct mechanisms. The annihilation
can produce a $Z^0$ boson that subsequently decays into the $f\bar f$
pair or they can be produced directly without the intermediate
production of a $Z^0$ boson. For this process it is not too difficult
to treat the reaction in terms of the stable external fermions. The
$Z^0$ has its 4-momentum fixed by the incoming $e^+e^-$ and therefore
does not participate in phase space integrals. Nevertheless, to account
for the Breit-Wigner resonance shape, some sort of resummation of the
perturbation expansion is required. Done carelessly and without due
regard to the constraints and requirements imposed by analytic $S$-matrix
theory, the resummation can lead to manifestly gauge-dependent results.
The solution to the problem is shown to be
to perform a Laurent expansion on the
complete matrix element about the pole. This naturally generates
exactly gauge-invariant expressions. An important point to note is that
the Laurent expansion is much more than a mathematical trick. There is
physics in the expansion. The leading term is the one responsible for the
Breit-Wigner resonance structure and describes the production, propagation
and subsequent decay of a physical unstable particle, in this case
the $Z^0$. The remaining non-resonant background accounts for the prompt
production of the final state fermions. As these processes are physically
distinguishable in principle it is essential that their corresponding
contributions to the matrix element be maintained separate and not combined
in any way. Thus a Laurent expansion should be performed even when using
a perturbation expansion, such as the background field
method\cite{DennWeigDitt}, that generates gauge-invariant Green functions.

The second process that will be considered is the process
$e^+e^-\rightarrow (f_1\bar f_1)(f_2\bar f_2)$ for energies above the
$Z^0Z^0$ production threshold. Here a complete treatment of the
matrix element in terms of the six external stable fermions becomes
unwieldy. The calculation would be simplified by considering just
$e^+e^-\rightarrow Z^0Z^0$, that is expected to be the dominant source
of $(f_1\bar f_1)(f_2\bar f_2)$, but $S$-matrix theory cannot tolerate
unstable particles as external states. Also the invariant mass of the
$Z^0$'s is not fixed but varies during phase-space integrations.
It will be shown, however, that from the matrix element for
$e^+e^-\rightarrow (f_1\bar f_1)(f_2\bar f_2)$, it is possible to
isolate a piece that describes propagation of $Z^0$'s over a finite-range.
This piece when suitably squared and summed over final state fermions
yields a physically meaningful cross-section for the process
$e^+e^-\rightarrow Z^0Z^0$. As such it must be exactly gauge-invariant.
It turns that the required part of the matrix element is precisely that
which is obtained by a Laurent expansion in the invariant masses of the
$Z^0$ bosons.

\section{Properties of Unstable Particles}

We begin by recalling what is known about unstable particles. $S$-matrix
theory tells us that unstable particles in intermediate states are
associated with poles in their invariant momentum, $s$,
in the $S$-matrix element lying off the physical
sheet below the real $s$-axis. The residue of the $S$-matrix element
factorizes as a consequence of Fredholm theory
(see ref.~\cite{ELOP}, p.~253). These residues can be used to define
generalized $S$-matrix elements for processes with unstable particles
as external states. These generalized $S$-matrix elements satisfy
unitarity relations that are analogous to those for stable particles but
continued off the real axis\cite{Stapp,Gunson}. Unfortunately the
unitarity relations no longer relate real quantities to one and other
and it is unclear what they have to do with physically measurable
cross-sections.

It is known that the lifetime of an unstable particle depends on how
it was prepared \cite{Schwinger}. The lifetime is therefore an
ill-defined concept without specifying further information such as that
the unstable particle is in a state of definite 4-momentum.

Veltman \cite{Veltman} showed that in models containing unstable particles
the $S$-matrix was unitary and causal on the Hilbert space spanned by
stable particle states. That is to say that only stable states must be
included in unitarity summations. Hence there is not really even room to
accommodate unstable particles as external states.

For an excellent review of some of the properties of unstable particles
see ref.~\cite{Martin}.

\section{Gauge Invariance near Resonance}

With the advent of LEP, the unstable $Z^0$ has been produced in large
numbers. The physics is described by the Standard Model Lagrangian
with which calculations can be done to arbitrary order in perturbation
theory. The catch is that the $Z^0$ resonance is a fundamentally
non-perturbative object and, in order to account for the resonance
structure, some sort of Dyson summation must be
performed. As first pointed out in ref.~\cite{Moriond}, the resummation
can lead to a breaking of gauge-invariance. To see how this happens
consider the process $e^+e^-\rightarrow f\bar f$ far from the
$Z^0$ resonance, say at PETRA energies of around 30\,GeV. There
perturbation theory works without difficulties as a resummation need not
and should not be performed. The results  generated by the perturbation
expansion are exactly gauge-invariant in each order. Let us look at
how the exact gauge-invariance is realized. In what follows we will use
$\Pi_{ZZ}^{(1)}(q^2)$ to denote the transverse part of the $Z^0$ boson
one-loop self-energy. The superscript in parentheses indicates the loop
order. The absence of a superscript will be taken to indicate the exact
expression to all orders and ${}^{(0)}$ used for vertices indicates the
tree-level. The initial state vertex correction that connects the initial-state
electron line to the $Z^0$ is given by $V_{iZ}(q^2)$ and the final-state
vertex connecting the $Z^0$ to the final-state fermion line by $V_{Zf}(q^2)$.
Although we will initially ignore the existence of the photon, the
transverse parts of the photon self-energy and the $Z$-$\gamma$
mixing are $\Pi_{\gamma\gamma}(q^2)$ and $\Pi_{Z\gamma}(q^2)$ and
respectively. $V_{i\gamma}(q^2)$ and $V_{\gamma f}(q^2)$ are the
initial- and final-state photon vertices. $B(s,t)$ denotes
one particle irreducible (1PI) corrections,
to the matrix element. These include things like as box diagrams.

For the process $e^+e^-\rightarrow f\bar f$ tree-level plus one-loop
corrections to the scattering amplitude read
\begin{equation}
A(s,t)=\frac{V_{iZ}^{(0)}V_{Zf}^{(0)}}{(s-M_Z^2)}
      +\frac{V_{iZ}^{(0)}\Pi_{ZZ}^{(1)}(s)V_{Zf}^{(0)}}{(s-M_Z^2)^2}
      +\frac{V_{iZ}^{(1)}(s)V_{Zf}^{(0)}}{(s-M_Z^2)}
      +\frac{V_{iZ}^{(0)}V_{Zf}^{(1)}(s)}{(s-M_Z^2)}+B^{(1)}(s,t)
\label{eq:PETRAcorr}
\end{equation}
that is exactly gauge-invariant. Here $M_Z$ is the unphysical renormalized
$Z^0$ boson mass. The correction (\ref{eq:PETRAcorr}) can be split into
separately gauge-invariant pieces by classifying the various contributions
according to their pole structure. Writing
\begin{eqnarray}
\frac{\Pi_{ZZ}^{(1)}(s)}{(s-M_Z^2)^2}&=&
     \frac{\Pi_{ZZ}^{(1)}(M_Z^2)}{ (s-M_Z^2)^2}
    +\frac{\Pi_{ZZ}^{(1)\prime}(M_Z^2)}{(s-M_Z^2)}
    +\Pi_{ZZ}^{(1)R}(s)\label{eq:sedecomp}\\
\frac{V_{iZ,Zf}^{(1)}(s)}{(s-M_Z^2)}&=&
     \frac{V_{iZ,Zf}^{(1)}(M_Z^2)}{ (s-M_Z^2)}
    +V_{iZ,Zf}^{(1)R}(s)\label{eq:vertexdecomp}
\end{eqnarray}
the coefficients of the pieces of (\ref{eq:PETRAcorr}) having a double,
single and no pole at $s=M_Z^2$ are
\begin{eqnarray}
&\Pi_{ZZ}^{(1)}(M_Z^2),&\label{eq:doublepole}\\
  &V_{iZ}^{(0)}\Pi_{ZZ}^{(1)\prime}(M_Z^2)V_{Zf}^{(0)}
   +V_{iZ}^{(1)}(M_Z^2)V_{Zf}^{(0)}+V_{iZ}^{(0)}V_{Zf}^{(1)}(M_Z^2),&
\label{eq:singlepole}\\
  &V_{iZ}^{(0)}\Pi_{ZZ}^{(1)R}(s)V_{Zf}^{(0)}
  +V_{iZ}^{(1)R}(s)V_{Zf}^{(0)}+V_{iZ}^{(0)}V_{Zf}^{(1)R}(s)
  +B^{(1)}(s,t),&\label{eq:nopole}
\end{eqnarray}
respectively. Since Eq.~(\ref{eq:PETRAcorr}) is exactly gauge-invariant
and since the cancellation of gauge-dependence cannot occur between
the various terms of differing pole structure we must conclude that
(\ref{eq:doublepole})--(\ref{eq:nopole}) are separately and exactly
gauge-invariant. We will find that these combinations will reappear
in the treatment of the $Z^0$ resonance. While Eq.~(\ref{eq:PETRAcorr})
is gauge-invariant it blows up hopelessly as $s\rightarrow M_Z^2$
and therefore cannot be used to treat the $Z^0$ resonance. What is
normally done is to perform a Dyson summation of the $Z^0$ self-energy
corrections to obtain an expression for the matrix element near
resonance
\begin{equation}
A(s,t)={V_{iZ}^{(0)}V_{Zf}^{(0)}
       +V_{iZ}^{(1)}(s)V_{Zf}^{(0)}
       +V_{iZ}^{(0)}V_{Zf}^{(1)}(s)\over s-M_Z^2-\Pi_{ZZ}^{(1)}(s)}
              +B^{(1)}(s,t).\label{eq:resummedwrong}
\end{equation}
The problem, as pointed out in ref.~\cite{Moriond}, is that $A(s,t)$
is now gauge-dependent. The reason is that the Dyson summation has
included all corrections of the form
$\left(\frac{\Pi_{ZZ}^{(1)}(s)}{s-M_Z^2}\right)^n$ each of which is
gauge-dependent. In the complete matrix element, the gauge-dependence
of these terms would be canceled by combinations of higher order
self-energy corrections, $\Pi_{ZZ}^{(n)}(s)$, vertex corrections,
$V_{iZ,Zf}^{(n)}(s)$, and 1PI corrections, $B^{(n)}(s,t)$.
None of these are present in Eq.~(\ref{eq:resummedwrong}) and hence it is
gauge-dependent. Far from resonance the gauge-dependence starts formally
at ${\cal O}(\alpha^2)$ compared to the tree-level result but near the
resonance where the leading $s-M_Z^2$ in resonant denominator of
Eq.~(\ref{eq:resummedwrong}) becomes small and the gauge-dependence
starts at ${\cal O}(\alpha)$ compared to the lowest order result.

What is the solution to this problem? One that has become popular is to
construct self-energy and vertex corrections that are gauge-invariant by
themselves so that the resummations cannot generate
spurious gauge-dependence\cite{KennedyLynn,KLIS,DegrassiSirlin}.
But this is only symptomatic relief for a much deeper illness which the
approach fails to address. In fact patching things in this way may generate
unforeseen and undesirable side-effects such as shifting the pole or
residue\cite{Stuart1}.

When $e^+e^-$ annihilate at energies near the $Z^0$ resonance they
produce a physical $Z^0$ that endures a while and then decays, normally
to a fermion-antifermion pair, $f\bar f$. In a {\it gedanken} world where
experimental resolution is extremely high or where the couplings of the
$Z^0$ are extremely weak, the presence of the $Z^0$ could be detected
as a physical particle in the way that neutrons, kaons and muons are.
When $e^+e^-$ annihilate they can also produce the final state
$f\bar f$ without producing a propagating $Z^0$. In this case no $Z^0$
would be detected no matter how good the experimental resolution or
how weak the couplings are made. Of course, there will always be a proportion
of $Z^0$ that will decay below the limit of experimental resolution.

The $f\bar f$ is produced by two distinguishable mechanisms and thus
the matrix element must always separate into the two corresponding
pieces. How can these two pieces be identified in the expression for
the matrix element? Consider the dressed propagator, for simplicity,
of an unstable scalar particle. In coordinate space it is
\begin{equation}
\Delta(x^\prime-x)=\int \frac{d^4k}{(2\pi)^4}
                   \frac{e^{-ik\cdot(x^\prime-x)}}
                                 {k^2-M^2-\Pi(k^2)+i\epsilon}
\label{eq:scalarpropagator}
\end{equation}
The integrand has a pole at $s=s_p$ that is a solution of the equation
$s_p-M^2-\Pi(s_p)=0$.
The denominator of the integrand may be written
\begin{equation}
k^2-M^2-\Pi(k^2)=\frac{k^2-s_p}{F(k^2)}\label{eq:Fdefn}
\end{equation}
The function $F(k^2)$ is analytic and
$F(s_p)=\left(1+\Pi^\prime(s_p)\right)^{-1}$.
Assuming $t\ne t^\prime$ and writing Eq.~(\ref{eq:scalarpropagator}) in the
form
\begin{equation}
\Delta(x^\prime-x)=\int\frac{d^4k}{(2\pi)^4} e^{-ik\cdot(x^\prime-x)}
 \left[\frac{F(s_p)}{k^2-s_p}+\frac{F(k^2)-F(s_p)}{k^2-s_p}\right].
\label{eq:splitprop}
\end{equation}
integration with respect to $k_0$ yields
\begin{equation}
\Delta(x^\prime-x)=-i\int \frac{d^3k}{(2\pi)^3 2k_0}
                   [e^{-ik\cdot(x^\prime-x)}\theta(t^\prime-t)
                   +e^{ik\cdot(x^\prime-x)}\theta(t-t^\prime)]F(s_p).
\label{eq:intsplitprop}
\end{equation}
The first term in Eq.~(\ref{eq:splitprop}) has split into two pieces and the
second has vanished because it has no poles. As both terms in
Eq.~(\ref{eq:splitprop}) are Lorentz invariant the vanishing of the second
in all reference frames for $t\ne t^\prime$ labels it as a contact
interaction. Similarly the first term is identifiable as an interaction
of finite space-time range. It has a pole at $k^2=s_p$ and is the leading
term in the Laurent expansion of the integrand about $s_p$.

This insight can now be applied to our generic process
$e^+e^-\rightarrow f\bar f$. The exact scattering amplitude to all
orders in perturbation theory takes the form
\begin{equation}
A(s,t)=\frac{V_{iZ}(s)V_{Zf}(s)}{s-M_Z^2-\Pi_{ZZ}(s)}+B(s,t).
\end{equation}
The right-hand side has a pole at $s=s_p$ and making a Laurent gives
\begin{equation}
A(s,t)=\frac{V_{iZ}(s_p)F_{ZZ}(s_p)V_{Zf}(s_p)}{s-s_p}
      +\frac{V_{iZ}(s)F_{ZZ}(s)V_{Zf}(s)
            -V_{iZ}(s_p)F_{ZZ}(s_p)V_{Zf}(s_p)}{s-s_p}
      +B(s,t).
\end{equation}
where $F_{ZZ}(s)$ is defined through the relation (\ref{eq:Fdefn}).
It may be shown \cite{Stuart2} that the pole position, residue and
background are separately and exactly gauge-invariant. They may
therefore be separately expanded as a perturbation series about $M_Z^2$
giving
\begin{eqnarray}
 A(s,t)&=&\frac{V_{iZ}^{(0)}V_{Zf}^{(0)}
        +V_{iZ}^{(0)}\Pi_{ZZ}^{(1)\prime}(M_Z^2)V_{Zf}^{(0)}
        +V_{iZ}^{(1)}(M_Z^2)V_{Zf}^{(0)}+V_{iZ}^{(0)}V_{Zf}^{(1)}(M_Z^2)
        }{s-s_p}\nonumber\\
     & &+V_{iZ}^{(0)}\Pi_{ZZ}^{(1)R}(s)V_{Zf}^{(0)}
        +V_{iZ}^{(1)R}(s)V_{Zf}^{(0)}+V_{iZ}^{(0)}V_{Zf}^{(1)R}(s)
              +B^{(1)}(s,t)\label{eq:summedright}
\end{eqnarray}
with, to ${\cal O}(\alpha^2)$,
\begin{equation}
s_p=M_Z^2+\Pi_{ZZ}^{(1)}(M_Z^2)+\Pi_{ZZ}^{(2)}(M_Z^2)
           +\Pi_{ZZ}^{(1)}(M_Z^2)\Pi_{ZZ}^{(1)\prime}(M_Z^2).
\end{equation}
Note that the pole position, residue and background are precisely the
gauge-invariant combinations identified in
(\ref{eq:doublepole})--(\ref{eq:nopole}). The overall result
Eq.~(\ref{eq:summedright}) is clearly gauge-invariant and naturally separates
into finite range and contact interaction as expected. As stated above
the resonant finite range piece is associated with the production of a
physical $Z^0$ and the contact interaction is associated with prompt
production of the final-state fermions via non-propagating modes of the
$Z^0$ field and box diagrams.

It should be emphasized that nowhere in this derivation did we introduce
an {\it ad hoc\/} width by hand as has been done by a number of
authors \cite{Moriond,NowaPila,BaurZeppenfeld,Papadopoulos}.
The finite width appeared naturally in the Laurent expansion about $s_p$.
The subsequent expansion of the pole position, residue and background
about the renormalized mass, $M_Z$, is justified because these quantities
represent three independent physical observables\cite{Stuart2}. Since the
procedure represents a well-defined sequence of expansions applied to the
complete matrix element, the result can be automatically guaranteed not
to violate unitarity or gauge-invariance.

Up to now the photon has been left out of the analysis. The Dyson summation
will clearly be complicated by the addition of photon exchange diagrams
and by $Z$-$\gamma$ mixing. Baulieu and Coquereaux \cite{BaulieuCoquereaux}
have shown how to perform the summation of the transverse parts exactly.
The result is that the full matrix element for $e^+e^-\rightarrow f\bar f$
is
\begin{eqnarray}
A(s,t)&=&V_{i\gamma}(s)\frac{s-M_Z^2-\Pi_{ZZ}(s)}
          {\left[s-\Pi_{\gamma\gamma}(s)\right]
           \left[s-M_Z^2-\Pi_{ZZ}(s)\right]-\Pi_{\gamma Z}^2(s)}
        V_{\gamma f}(s)\nonumber\\
     &+&V_{i\gamma}(s)\frac{\Pi_{\gamma Z}(s)}
          {\left[s-\Pi_{\gamma\gamma}(s)\right]
           \left[s-M_Z^2-\Pi_{ZZ}(s)\right]-\Pi_{\gamma Z}^2(s)}
        V_{Zf}(s)\nonumber\\
     &+&V_{iZ}(s)\frac{\Pi_{Z\gamma}(s)}
          {\left[s-\Pi_{\gamma\gamma}(s)\right]
           \left[s-M_Z^2-\Pi_{ZZ}(s)\right]-\Pi_{\gamma Z}^2(s)}
        V_{\gamma f}(s)\nonumber\\
     &+&V_{iZ}(s)\frac{s-\Pi_{\gamma\gamma}(s)}
          {\left[s-\Pi_{\gamma\gamma}(s)\right]
           \left[s-M_Z^2-\Pi_{ZZ}(s)\right]-\Pi_{\gamma Z}^2(s)}
        V_{Zf}(s)\nonumber\\
     &+&B(s,t)\label{eq:BCresum}
\end{eqnarray}
exactly.
The first term, representing photon exchange, can be split into a resonant
and non-resonant piece. Collecting the resonant pieces together yields
\begin{eqnarray}
A(s,t)&=&\frac{\left[V_{i\gamma}(s)
               \frac{\displaystyle\Pi_{\gamma Z}(s)}
                    {\displaystyle s-\Pi_{\gamma\gamma}(s)}
               +V_{iZ}(s)\right]
         \left[V_{Zf}(s)
        +\frac{\displaystyle\Pi_{Z\gamma}(s)}
              {\displaystyle s-\Pi_{\gamma\gamma}(s)}
         V_{\gamma f}(s)\right]}
              {s-M_0^2-\Pi_{ZZ}(s)
               -\frac{\displaystyle\Pi_{\gamma Z}^2(s)}
                     {\displaystyle s-\Pi_{\gamma\gamma}(s)}
               }\nonumber\\
& &\qquad+\frac{V_{i\gamma}(s)V_{\gamma f}(s)}{s-\Pi_{\gamma\gamma}(s)}
       +B(s,t).\label{eq:ressumm}
\end{eqnarray}

This exact scattering amplitude has a pole at the point, $s_p$, satisfying
the equation
\begin{equation}
s_p-M_Z^2-\Pi_{ZZ}(s_p)
-\frac{\Pi_{Z\gamma}^2(s_p)}{s_p-\Pi_{\gamma\gamma}(s_p)}=0.
\end{equation}
Defining the function $F_{ZZ}(s_p)$ from the relation
\begin{equation}
s-M_Z^2-\Pi_{ZZ}(s)
-{\Pi_{Z\gamma}^2(s)\over s-\Pi_{\gamma\gamma}(s)}={1\over F_{ZZ}(s)}(s-s_p)
\end{equation}
and extracting the leading term of Eq.~(\ref{eq:ressumm})
in the Laurent expansion about $s_p$ leads to
\begin{eqnarray}
A(s,t)&=&{R_{iZ}(s_p)R_{Zf}(s_p)\over s-s_p}\nonumber\\
      &+&{R_{iZ}(s)R_{Zf}(s)-R_{iZ}(s_p)R_{Zf}(s_p)\over s-s_p}
      +{V_{i\gamma}(s)V_{\gamma f}(s)\over s-\Pi_{\gamma\gamma}(s)}
      +B(s,t)
\end{eqnarray}
in which
\begin{eqnarray}
R_{iZ}(s)&=&\left[V_{i\gamma}(s)
   \frac{\Pi_{\gamma Z}(s)}{s-\Pi_{\gamma\gamma}(s)}
                     +V_{iZ}(s)\right]F_{ZZ}^\frac{1}{2}(s),
\nonumber\\
R_{Zf}(s)&=&F_{ZZ}^\frac{1}{2}(s)\left[V_{Zf}(s)
       +\frac{\Pi_{Z\gamma}(s)}{s-\Pi_{\gamma\gamma}(s)}
                V_{\gamma f}(s)\right].
\nonumber
\end{eqnarray}

The equation (\ref{eq:ressumm}) is {\bf exact}. It is remarkably simple in
structure; much more so than many approximate formulas that have appeared
in the literature. The factorization of the residue at the pole, demanded
by analytic $S$-matrix theory is manifest. The photon exchange contribution
appears in the background in a transparent form. As with
Eq.~(\ref{eq:summedright}) the pole position, residue and background can
be expanded separately as well-behaved perturbation expansions about
the renormalized mass $M_Z^2$. The background is regular in $s$ and can
be expanded as a Taylor series. The Laurent expansion has separated the
finite range and contact interaction contributions to the matrix element
and should be performed even in the case where gauge-invariant
self-energies are produced in a consistent manner such as with the
background field method\cite{DennWeigDitt}. Expanding in the way shown here,
however, makes the necessity or advantage of gauge-invariant self-energies
unclear.

The results obtained in this section have been shown to be consistent
with Ward identities\cite{HVeltman}.

\section{The Mass and Width of an Unstable Particle}

The quantity $M_Z$ in the formulas given in the previous section is the
renormalized mass of the $Z^0$ boson in the scheme dictated by the
counterterms contained in the self-energies and vertex corrections.
Expressions for the counterterms appearing in the pole position,
residue and background in the Standard Model for a general renormalization
scheme can be found in ref.~\cite{Stuart2}.

The renormalized mass is merely a bookkeeping device that has no physical
content. This is clear from the fact that, in the $\overline{\rm MS}$
renormalization scheme, the renormalized mass depends on the arbitrary
scale, $\mu$, and in the on-shell scheme it is
gauge-dependent\cite{SirlinMass1,SirlinMass2} as it is
entitled to be. The physical quantity associated with any particle is
the position of the pole of its dressed propagator. For a stable
particle the pole position is real and it may be identified with the
physical mass of the particle. The renormalized mass may be set
equal to this physical mass by appropriate choice of counterterms.

For an unstable particle the pole position is complex. It is the pole
position as a whole that is the physically meaningful entity. The real
and imaginary parts taken separately have no physical significance as
they never appear separately in any expression for a physical observable.
They are no more meaningful than, say, the modulus and argument of the
pole position. Traditionally\cite{Peierls,Levy} for convenience the pole
was decomposed in two possible ways,
\begin{equation}
s_p=M_Z^2-i\Gamma_Z M_Z
\label{eq:massdef1}
\end{equation}
or
\begin{equation}
s_p=\left(M_Z-i\frac{\Gamma_Z}{2}\right)^2.
\label{eq:massdef2}
\end{equation}
Both definitions are arbitrary. It was pointed out independently in
ref.s~\cite{Stuart1} and \cite{WillenbrockValencia} that the traditional
definition of the mass lies below the on-shell renormalized mass,
ostensibly being extracted by LEP, by 34\,MeV in the case of the definition
(\ref{eq:massdef1}) and 26\,MeV in the case of (\ref{eq:massdef2}).
Sirlin\cite{SirlinMass1} modified the definition of the renormalized mass
in the on-shell scheme to be just $(M_Z^{\rm OS})^2=M_Z^2-\Gamma_Z^2$ with
$M_Z$ and $\Gamma_Z$ being taken from (\ref{eq:massdef1}). It is not clear
however that this can play the r\^ole of a self-consistent renormalized
mass without violating Ward identities. Further discussion of these
points can be found in ref.~\cite{Stuart3} where it is also suggested that
the residue factors at the pole may be used to define model-independent
partial widths.

At this point a word is appropriate about the range of validity of the
Laurent expansion used in the previous section.
It is correct up to a radius determined by the nearest
singularity to the pole, $s_p$. If there are thresholds in the resonance
region then these originate branch cuts and the expansion breaks down.
However some interesting physics then appears. In fact for the $Z^0$,
there are a multitude
thresholds separated, at most by the mass of a light fermion pair, lying
under the resonance. For example, $Z^0\rightarrow W^+b\bar bs\bar c$,
$Z^0\rightarrow 10(b\bar b)$ lie under the umbrella resonance but
are sufficiently
weak as make them entirely negligible. Suppose for a moment that the top
quark had had a mass that was $m_t\approx M_Z/2$ then the first-order
$t\bar t$ threshold would lie under the peak of the
resonance. The resonance region would be
under the influence of two distinct poles, one reached by crossing the
real $s$-axis onto the unphysical sheet below the threshold branch cut
and the other by crossing above.
Bhattacharya and Willenbrock\cite{BhattaWillenbrock} have examined
this scenario in a toy model and found that the resonance peak becomes
asymmetrical being narrower on the low energy side which is easily
understood in that the decay width increases as the threshold is
crossed. In the situation where $m_t\approx M_Z/2$ both poles
must be associated with the $Z^0$ boson. The $Z^0$ resonance then
becomes a closely-spaced doublet with a separation of roughly 100\,MeV
\cite{Stuart3}.

\section{Production Cross-sections for Unstable Particles}

In the foregoing we have treated the complete process
$e^+e^-\rightarrow f\bar f$. This process possesses a number of simplifying
features. There are only a total of four external particles and the
4-momentum of the intermediate unstable $Z^0$ is fixed by the incoming
$e^+e^-$. In contrast the process
$e^+e^-\rightarrow(f_1\bar f_1)(f_2\bar f_2)$
has an total of  six external particles and therefore is much more
complicated to deal with especially if radiative corrections are to be
computed. It would be advantageous if the process $e^+e^-\rightarrow Z^0Z^0$
could be treated in some meaningful way as it is expected to be the dominant
mechanism operating in the production of the $(f_1\bar f_1)(f_2\bar f_2)$.
Now instead of single unstable particle there are two and their 4-momenta
are are no longer fixed but must be integrated over in phase-space
integrations. The blind application of kinematic identities can generate
complex scattering angles because of the presence of imaginary parts
in particle self-energies. Some care is therefore required.

As stated earlier $S$-matrix theory cannot treat unstable particles as
asymptotic states but as shown above it is possible to isolate
the pieces of the matrix element for
$e^+e^-\rightarrow(f_1\bar f_1)(f_2\bar f_2)$
that correspond to the production,
propagation and subsequent decay of physical $Z^0$ bosons by means of
the Laurent expansion. Summing over all possible stable decay products
then gives the total production cross-section for the unstable particle.

Considering $e^+e^-\rightarrow Z^0Z^0$ avoids the need to confront the
the issue of the Coulomb singularity that is present in
$e^+e^-\rightarrow W^+W^-$ when the $W^+W^-$ pair are produced at low
relative velocity\cite{FadKhoMar,BardBeenDenn,FadKhoMarCha}.
At lowest order there
are no Feynman diagrams containing the triple vector boson vertex.

With $u$ and $v$, as usual, used to denote the spinor wavefunctions of
the external fermions, the part of the full matrix element that can generate
finite-range interactions is
\begin{eqnarray}
{\cal M}=\sum_i [\bar v_{e^+} T^i_{\mu\nu} u_{e^-}]
M_i(s,t,u,p_1^2,p_2^2)
&\times&\frac{1}{p_1^2-M_Z^2-\Pi_{ZZ}^2(p_1^2)}
[\bar u_{f_1} V_{Zf_1}^\mu(p_1^2) v_{\bar f_1}]\nonumber\\
&\times&\frac{1}{p^2_2-M_Z^2-\Pi_{ZZ}^2(p_2^2)}
[\bar u_{f_2}V_{Zf_2}^\nu(p_2^2) v_{\bar f_2}]
\label{eq:fullZZ}
\end{eqnarray}
where $T_{\mu\nu}^i$ are kinematic tensors that span the tensor structure
of the matrix element and the $M_i$ are the associated form factors that
are analytic in their arguments. It may be convenient, although not
at all necessary, to construct the $T_{\mu\nu}^i$ so as to be individually
gauge-invariant \cite{BardeenTung}. $M_i$ is a function of the
usual Mandelstam variables $s$, $t$ and $u$ and of the invariant masses
$p_1^2$ of the $f_1\bar f_1$ pair and $p_2^2$ of the $f_2\bar f_2$.

The expression (\ref{eq:fullZZ}) has been obtained by Dyson summation of
the $Z^0$ self-energies. It is not by itself gauge-invariant as it still
contains contact interaction terms that must be split off. In fact the
complete matrix element for $e^+e^-\rightarrow(f_1\bar f_1)(f_2\bar f_2)$
naturally separates into four distinguishable processes according to
whether not it is resonant or not in $p_1^2$ or $p_2^2$. Fredholm
theory guarantees the exact factorization of final-state contributions
in the resonant channels. The four distinguishable physical processes are
\[
\begin{array}{cl}
e^+e^-\longrightarrow&\left\{ \begin{array}{l}
           Z^0\rightarrow f_1\bar f_1\\
           Z^0\rightarrow f_2\bar f_2
           \end{array}\right.\\
 & \\
e^+e^-\longrightarrow&\left\{ \begin{array}{l}
           Z^0\rightarrow f_1\bar f_1\\
           f_2\bar f_2
           \end{array}\right.\\
 & \\
e^+e^-\longrightarrow&\left\{ \begin{array}{l}
           f_1\bar f_1\\
           Z^0\rightarrow f_2\bar f_2
           \end{array}\right.\\
 & \\
e^+e^-\longrightarrow&\left\{ \begin{array}{l}
           f_1\bar f_1\\
           f_2\bar f_2
           \end{array}\right.
\end{array}\]
As for $e^+e^-\rightarrow f\bar f$ they become physically distinguishable in
the {\it gedanken\/} world of very high detector resolution or very weak
couplings.

As it stands Eq.~(\ref{eq:fullZZ}) contains the entire contribution
of the matrix element for the
first process listed above and parts of the remaining three. Additional
Feynman diagrams must be included to compute the full matrix elements
for these last three.

Once again the $Z$-$\gamma$ mixing has been neglected but it is a simple
matter to include it in the manner described previously.

Performing the Laurent expansion in $p_1^2$ and $p_2^2$ doubly
resonant part corresponding to the process
$e^+e^-\rightarrow(Z^0\rightarrow f_1\bar f_1)(Z^0\rightarrow f_2\bar f_2)$
is
\begin{eqnarray}
{\cal M}=\sum_i [\bar v_{e^+} T^i_{\mu\nu} u_{e^-}]
M_i(s,t,u,s_p,s_p)
&\times&\frac{F_{ZZ}(s_p)}{p_1^2-s_p}
[\bar u_{f_1}V_{Zf_1}^\mu(s_p) v_{\bar f_1}]\nonumber\\
&\times&\frac{F_{ZZ}(s_p)}{p_2^2-s_p}
[\bar u_{f_2}V_{Zf_2}^\mu(s_p) v_{\bar f_2}]
\label{eq:resZZ}
\end{eqnarray}
in which the final-state fermion currents factorize.
Because Eq.~(\ref{eq:resZZ})
is the doubly resonant part of the complete matrix element it must be
gauge-invariant. In lowest it order it is
\begin{equation}
{\cal M}=\sum_{i=1}^2 [\bar v_{e^+} T^i_{\mu\nu} u_{e^-}]
M_i.\frac{1}{p_1^2-s_p}
[\bar u_{f_1}V_{Zf_1}^\mu v_{\bar f_1}].\frac{1}{p_2^2-s_p}
[\bar u_{f_2}V_{Zf_2}^\mu v_{\bar f_2}]
\label{eq:lowestZZ}
\end{equation}
in which
$T^1_{\mu\nu}M_1=\gamma_\mu(\slash{p}_{e^-}-\slash{p}_1)\gamma_\nu/t$,
$T^2_{\mu\nu}M_2=\gamma_\nu(\slash{p}_{e^-}-\slash{p}_2)\gamma_\mu/u$.
The final state vertices then take the form
$V_{Zf}^\mu=ie\gamma_\mu(\beta_L^f\gamma_L+\beta^f_R\gamma_R)$.
As usual $\gamma_L$ and $\gamma_R$ are the left- and right-hand
helicity projection operators. The left- and right-handed couplings
of the $Z^0$ to a fermion $f$ are
\[
\beta_L^f=\frac{t_3^f-\sin^2\theta_W Q^f}{\sin\theta_W\cos\theta_W},
\hbox to 2cm{}
\beta_R^f=-\frac{\sin\theta_W Q^f}{\cos\theta_W}.
\]
Squaring the matrix element and integrating over the final-state
fermion momenta but with fixed invariant mass $p_1^2$ and $p_2^2$
gives the differential cross-section for physical $Z^0$ production
\begin{equation}
\frac{\partial^3\sigma}{\partial t\,\partial p_1^2\,\partial p_2^2}
                   =\frac{\pi\alpha^2}{s^2}
                    (\vert\beta_L^e\vert^4+\vert\beta_R^e\vert^4)
                     \left\{\frac{t}{u}+\frac{u}{t}
                     +\frac{2(p_1^2+p_2^2)}{ut}
-p_1^2p_2^2\left(\frac{1}{t^2}+\frac{1}{u^2}\right)\right\}
\ \rho(p_1)\ \rho(p_2)
\end{equation}
with
\begin{eqnarray*}
\rho(p)&=&\frac{\alpha}{6\pi}
    \sum_f(\vert\beta_L^f\vert^2+\vert\beta_R^f\vert^2)
          \frac{p^2}{\vert p^2-s_p\vert^2}
               \theta(p_0)\theta(p^2)\\
 &\approx&\frac{1}{\pi}
         .\frac{p^2 (\Gamma_Z/M_Z)}{(p^2-M_Z^2)^2+\Gamma_Z^2 M_Z^2}
          \theta(p_0)\theta(p^2)
\end{eqnarray*}
The $\theta$ functions are generated in the phase space integrations
and select the positive energy component. Thus the differential cross-section
decribes the production of an object that has both finite range
and positive energy as would be expected for a physical particle.
The function $\rho(p)\rightarrow \delta(p^2-M_Z^2)\theta(p_0)$
as $\Im(s_p)\rightarrow 0$ that is the well-known result obtained by cutting
in free propagator. Final state interactions could be built in to cross-section
via the convolution kernel $\rho(p_1).\rho(p_2)$.

Integrating over $t$, $p_1^2$ and $p_2^2$ gives the total production
cross-section for $e^+e^-\rightarrow Z^0Z^0$ gives
\begin{equation}
\sigma(s)=\int_0^s dp_1^2
          \int_0^{(\sqrt{s}-\sqrt{p_1^2})^2} dp_2^2
          \sigma(s;p_1^2,p_2^2)\ \rho(p_1^2)\ \rho(p_2^2),
\label{eq:ZZxsec}
\end{equation}
where
\[
\sigma(s;p_1^2,p_2^2)=\frac{2\pi\alpha^2}{s^2}
       (\vert\beta_L^e\vert^4+\vert\beta_R^e\vert^4)
       \left\{\left(\frac{1+(p_1^2+p_2^2)^2/s^2}
{1-(p_1^2+p_2^2)/s}\right)\ln\left(\frac{-s+p_1^2+p_2^2+\lambda}
                                    {-s+p_1^2+p_2^2-\lambda}\right)
-\frac{\lambda}{s}\right\}
\]
and $\lambda=\sqrt{s^2+p_1^4+p_2^4-2sp_1^2-2sp_2^2-2p_1^2p_2^2}$.
Setting $p_1^2=p_2^2=M_Z^2$ reproduces the results of Brown and
Mikaelian\cite{Brown}.

The result given here is superficially rather similar to those of
other authors\cite{MutaNajiWaka,DennSack}. The difference lies in the
fact in the present case strictly only the doubly resonant part of the
matrix element has been included and manifests itself in a change in
the convolution kernel $\rho$. In higher orders the approach of
ref.s~\cite{MutaNajiWaka} and \cite{DennSack} will lead to the evaluation
of off-shell matrix elements that will generally be gauge-dependent.
In the method described here the form factors in the matrix element are
evaluated at the pole and are therefore gauge-invariant. This also
simplifies the evaluation of the integrals like those in
Eq.~(\ref{eq:ZZxsec}) as the
higher-order form-factors are not integrated over. Only factors deriving
from the kinematic tensors are integrated.

After the $Z^0$'s have propagated and decayed the final state fermions
so-produced can interact. Because the $Z^0$'s have propagated a finite distance
from their point of production, the final-state interactions are expected
to be suppressed. This is generally found to be the
case\cite{MelnYako1,MelnYako2,Khoze}. Such a suppression was already
observed\cite{KuhnStuart,KuhnJadaStuaWas} in
the case of the $e^+e^-\rightarrow Z^0\rightarrow f\bar f$
where the exact correction to the cross-section from the interference of
initial- and final-state interference was shown to vanish near resonance
and may be understood as suppression due the finite range propagated by
the $Z^0$.

\section{Conclusions}

In the foregoing paper it has been shown how to use the Laurent expansion
to generate finite-order matrix elements, for processes involving
unstable particles, that are exactly
gauge-invariant. This is achieved without the {\it ad hoc\/} introduction
of finite widths and without the need for gauge-invariant self-energies
or vertex corrections. Indeed, because of the interpretation of the
resonant part as being the finite range interaction of an unstable
particle being produced, propagating and decaying, the Laurent expansion
should be carried out in all calculational schemes in order to preserve
the separation of the matrix element into physically distinguishable parts.

The approach was also used in obtaining physically meaningful expressions
for the production cross-sections for the production cross-sections for
unstable particles.


\begin{thebibliography}{99}

\bibitem{Pilaftsis} A. Pilaftsis,
                   {\it Z.\ Phys.}\ {\bf C 47} (1990) 95.

\bibitem{EilaHeweSoni} G. Eilam, J. L. Hewett and A. Soni,
                      {\it Phys. Rev. Lett.}\ {\bf 67} (1991) 1979.

\bibitem{Moriond} R. G. Stuart, in {\it $Z^0$ Physics; Proceedings of the
XXVth Rencontre de Moriond}, ed.\ J.~Tran Thanh Van, Editions Fronti\'eres,
Gif-sur-Yvette (1990) p.41

\bibitem{NowaPila} M. Nowakowski and A. Pilaftsis,
                   {\it Z.\ Phys.}\ {\bf C 60} (1993) 121.



\bibitem{BaurZeppenfeld} U. Baur and D. Zeppenfeld,
                   Madison preprint MAD/PH/878,
                   Bulletin Board hep-ph 9503304.

\bibitem{Papadopoulos} C. G. Papadopoulos, CERN preprint CERN-TH/95-46,
                   Bulletin Board hep-ph 9503276.

\bibitem{AeppCuypOlde} A. Aeppli, F. Cuypers and G. J. van Oldenborgh,
                       {\it Phys.\ Lett.}\ {\bf B 314} (1993) 413.

\bibitem{AeppOldeWyl} A. Aeppli, G. J. van Oldenborgh and D. Wyler,
                      {\it Nucl.\ Phys.}\ {\bf B 428} (1994) 126.

\bibitem{Stuart1} R. G. Stuart, {\it Phys.\ Lett.}\ {\bf B 262} (1991) 113.

\bibitem{Stuart2} R. G. Stuart, {\it Phys.\ Lett.}\ {\bf B 272} (1991) 353.

\bibitem{Stuart3} R. G. Stuart, {\it Phys. Rev. Lett.}\ {\bf 70} (1993) 3193.

\bibitem{Stuart4} R. G. Stuart, Michigan preprint UM-TH-95-06,
                   Bulletin Board hep-ph 9504215.

\bibitem{DennWeigDitt} A. Denner, G. Weiglein and S. Dittmaier,
                       {\it Phys.\ Lett.}\ {\bf B 333} (1994) 420;
                       Bielefeld preprint BI-TP-94-50, Bulletin Board
                       hep-ph 9410338.

\bibitem{ELOP} R. J. Eden, P. V. Landshoff, D. I. Olive
                  and J. C. Polkinghorne,
                  {\it The Analytic $S$-Matrix}, Cambridge University
                  Press, Cambridge (1966).

\bibitem{Stapp} H. P. Stapp, {\it Nuovo Cimento} {\bf 32} (1964) 103.

\bibitem{Gunson} J. Gunson, {\it J.\ Math.\ Phys.}\ {\bf 6} (1965) 827;
                            {\bf 6} (1965) 845; {\bf 6} (1965) 852.

\bibitem{Schwinger} J. Schwinger, {\it Ann.\ Phys.}\ {\bf 9} (1960) 169.

\bibitem{Veltman} M. Veltman, {\it Physica\/} {\bf 29} (1963) 186.

\bibitem{Martin} A. Martin, in {\it $Z^0$ Physics 1990; NATO Advanced Study
                 Institute Cargese Summer School\/} NATO ASI (1990) p.483.

\bibitem{KennedyLynn} D. C. Kennedy and B. W. Lynn,
                      {\it Nucl.\ Phys.}\ {\bf B 322} (1989) 1.

\bibitem{KLIS} D. C. Kennedy, B. W. Lynn, C. J.-C. Im and R. G. Stuart,
                      {\it Nucl.\ Phys.}\ {\bf B 321} (1989) 83.

\bibitem{DegrassiSirlin} G. Degrassi and A. Sirlin,
                         {\it Phys.\ Rev.}\ {\bf D 46} (1992) 3104.

\bibitem{BaulieuCoquereaux} L. Baulieu and R. Coquereaux,
                      {\it Ann.\ Phys.}\ {\bf 140} (1982) 163

\bibitem{HVeltman} H. Veltman, {\it Z.\ Phys.}\ {\bf C 62} (1994) 35.

\bibitem{SirlinMass1} A. Sirlin, {\it Phys. Rev. Lett.}\ {\bf 67} (1991) 2127.

\bibitem{SirlinMass2} A. Sirlin, {\it Phys.\ Lett.}\ {\bf B 267} (1991) 240.

\bibitem{Peierls} R. E. Peierls, {\it Proceedings of the 1954 Glasgow
                                 Conference on Nuclear and Meson Physics},
                                 Pergamon Press, New York, (1955) 296.

\bibitem{Levy} M. L\'evy, {\it Nuovo Cimento} {\bf 13} (1959) 115.

\bibitem{WillenbrockValencia} S. Willenbrock and G. Valencia,
                              {\it Phys.\ Lett.}\ {\bf B 259} (1991) 373.

\bibitem{BhattaWillenbrock} T. Bhattacharya and S. Willenbrock,
                Brookhaven preprint, BNL-56481

\bibitem{FadKhoMar} V. S. Fadin, V. A. Khoze and A. D. Martin,
                       {\it Phys.\ Lett.}\ {\bf B 311} (1993) 403.

\bibitem{BardBeenDenn} D. Bardin, W. Beenakker and A. Denner,
                       {\it Phys.\ Lett.}\ {\bf B 317} (1993) 213.

\bibitem{FadKhoMarCha} V. S. Fadin, V. A. Khoze, A. D. Martin and A. Chapovsky,
            Durham preprint DTP/94/116, Bulletin board hep-ph 9501214.

\bibitem{BardeenTung} W. A. Bardeen and W.-K. Tung,
                         {\it Phys.\ Rev.}\ {\bf 173} (1968) 1423.

\bibitem{Brown} R. W. Brown and K. O. Mikaelian,
                         {\it Phys.\ Rev.}\ {\bf D 19} (1979) 922.

\bibitem{MutaNajiWaka} T. Muta, R. Najima and S. Wakaizumi,
                       {\it Mod.\ Phys.\ Lett.}\ {\bf A1} (1986) 203.

\bibitem{DennSack} A. Denner and T. Sack,
                   {\it Z.\ Phys.}\ {\bf C 45} (1990) 439.

\bibitem{MelnYako1} K. Melnikov and O. Yakovlev,
                       {\it Phys.\ Lett.}\ {\bf B 324} (1994) 217;

\bibitem{MelnYako2} K. Melnikov and O. Yakovlev, Mainz preprint MZ-TH/95-01,
                   Bulletin Board hep-ph 9501358.

\bibitem{Khoze} V. A. Khoze, Durham preprint DTP-94-114, Bulletin Board
                             hep-ph 9412239.

\bibitem{KuhnStuart} J. H. K\"uhn and R. G. Stuart,
                       {\it Phys.\ Lett.}\ {\bf B 200} (1988) 360;

\bibitem{KuhnJadaStuaWas} J. H. K\"uhn, S. Jadach, R. G. Stuart and Z. W\c as,
                   {\it Z.\ Phys.}\ {\bf C 38} (1988) 609;
                     E {\it ibid.}\ {\bf C 45} (1990) 528;



\end{thebibliography}
\end{document}